\documentstyle[epsf,referee]{mn}
\newif\ifAMStwofonts
\AMStwofontstrue

\ifoldfss
  \ifCUPmtlplainloaded \else
    \NewTextAlphabet{textbfit} {cmbxti10} {}
    \NewTextAlphabet{textbfss} {cmssbx10} {}
    \NewMathAlphabet{mathbfit} {cmbxti10} {} % for math mode
    \NewMathAlphabet{mathbfss} {cmssbx10} {} %  "   "    "
  \fi
  \ifAMStwofonts
    \ifCUPmtlplainloaded \else
      \NewSymbolFont{upmath} {eurm10}
      \NewSymbolFont{AMSa} {msam10}
      \NewMathSymbol{\upi}     {0}{upmath}{19}
      \NewMathSymbol{\umu}     {0}{upmath}{16}
      \NewMathSymbol{\upartial}{0}{upmath}{40}
      \NewMathSymbol{\leqslant}{3}{AMSa}{36}
      \NewMathSymbol{\geqslant}{3}{AMSa}{3E}

      \let\leq=\leqslant \let\le=\leqslant
      \let\geq=\geqslant 
    \fi
  \fi
\fi % End of OFSS

\ifnfssone
  \newmathalphabet{\mathit}
  \addtoversion{normal}{\mathit}{cmr}{m}{it}
  \addtoversion{bold}{\mathit}{cmr}{bx}{it}
  \newmathalphabet{\mathbfit} % math mode version of \textbfit{..}
  \addtoversion{normal}{\mathbfit}{cmr}{bx}{it}
  \addtoversion{bold}{\mathbfit}{cmr}{bx}{it}
  \newmathalphabet{\mathbfss} % math mode version of \textbfss{..}
  \addtoversion{normal}{\mathbfss}{cmss}{bx}{n}
  \addtoversion{bold}{\mathbfss}{cmss}{bx}{n}
  \ifAMStwofonts
    \ifCUPmtlplainloaded \else
      \UseAMStwoboldmath
      \makeatletter
      \new@mathgroup\upmath@group
      \define@mathgroup\mv@normal\upmath@group{eur}{m}{n}
      \define@mathgroup\mv@bold\upmath@group{eur}{b}{n}
      \edef\UPM{\hexnumber\upmath@group}
      \new@mathgroup\amsa@group
      \define@mathgroup\mv@normal\amsa@group{msa}{m}{n}
      \define@mathgroup\mv@bold\amsa@group{msa}{m}{n}
      \edef\AMSa{\hexnumber\amsa@group}
      \makeatother
      \mathchardef\upi="0\UPM19
      \mathchardef\umu="0\UPM16
      \mathchardef\upartial="0\UPM40
      \mathchardef\leqslant="3\AMSa36
      \mathchardef\geqslant="3\AMSa3E

      \let\leq=\leqslant \let\le=\leqslant
      \let\geq=\geqslant 
    \fi
  \fi
\fi % End of NFSS release 1

\ifnfsstwo
  \DeclareMathAlphabet{\mathbfit}{OT1}{cmr}{bx}{it}
  \SetMathAlphabet\mathbfit{bold}{OT1}{cmr}{bx}{it}
  \DeclareMathAlphabet{\mathbfss}{OT1}{cmss}{bx}{n}
  \SetMathAlphabet\mathbfss{bold}{OT1}{cmss}{bx}{n}
  \ifAMStwofonts
    \ifCUPmtlplainloaded \else
      \DeclareSymbolFont{UPM}{U}{eur}{m}{n}
      \SetSymbolFont{UPM}{bold}{U}{eur}{b}{n}
      \DeclareSymbolFont{AMSa}{U}{msa}{m}{n}
      \DeclareMathSymbol{\upi}{0}{UPM}{"19}
      \DeclareMathSymbol{\umu}{0}{UPM}{"16}
      \DeclareMathSymbol{\upartial}{0}{UPM}{"40}
      \DeclareMathSymbol{\leqslant}{3}{AMSa}{"36}
      \DeclareMathSymbol{\geqslant}{3}{AMSa}{"3E}

      \let\leq=\leqslant \let\le=\leqslant
      \let\geq=\geqslant 
    \fi
  \fi
\fi % End of NFSS release 2

\ifCUPmtlplainloaded \else
  \ifAMStwofonts \else % If no AMS fonts
    \def\upi{\pi}
    \def\umu{\mu}
    \def\upartial{\partial}
  \fi
\fi
\title{Higher Criticism Statistic: Detecting and Identifying Non-Gaussianity in the WMAP First Year Data}
\author[] {
L. Cay\'on$^{1}$, J. Jin$^2$ and A. Treaster$^{1}$ \\
1. Department of Physics. Purdue University. 525 Northwestern Avenue, West Lafayette, IN 47907-2036.\\
2. Department of Statistics. Purdue University. 150 N. University Street, West Lafayette, IN 47907-2067.\\}

\date{\today}

\pagerange{\pageref{firstpage}--\pageref{lastpage}}
\pubyear{2001}

\begin{document}

\maketitle

\label{firstpage}

\begin{abstract}

\noindent  Higher Criticism is a recently developed statistic for non-Gaussian detection, 
proposed in Donoho $\&$ Jin 2004, where it has been shown to be 
effective at resolving a very subtle
testing problem: whether $n$ normal means are all zero versus 
the alternative that a small fraction is nonzero.  Higher Criticism is 
also useful in the detection of a non-Gaussian convolution component of 
cosmic strings in the Cosmic Microwave Background (CMB), see Jin et al. 
2004.  In this paper, we study how well the anisotropies 
of the CMB fit with the homogeneous and isotropic Gaussian distribution 
predicted by the Standard Inflationary model.  
We find that Higher Criticism is useful for two purposes.
First, Higher Criticism has competitive detection power, and 
non-Gaussianity is detected at the level $99\%$ in the first year WMAP data.
We generated $5000$ Monte Carlo Gaussian
simulations of the CMB maps. By applying the Higher Criticism to all 
of these maps in wavelet space, we constructed confidence regions of Higher 
Criticism at levels $68\%$, $95\%$, and $99\%$.  We find that the Higher 
Criticism value of WMAP is outside the $99\%$ confidence region at 
a wavelet scale of $5$ degrees ($99.46\%$ of Higher Criticism values 
based on simulated maps are below the values for WMAP).
Second, Higher Criticism offers a way to locate a 
small portion of data that accounts for the non-Gaussianity. This property is 
not immediately available for other statistical tests such as the widely-used excess kurtosis test.
Using Higher Criticism, we have successfully identified a ring of pixels centered 
at $(l\approx 209^{\circ}, b\approx -57^{\circ})$, which seems to account for 
the observed detection of non-Gaussianity at the wavelet scale of 5 degrees. By removing the ring from the WMAP 
data set, no more prominent deviation from Gaussianity was found. 
Note that the detection is achieved in wavelet space first. Second, it is
always possible that a fraction of
pixels within the ring might deviate from Gaussianity even if they do not 
appear to be above the $99\%$ confidence level in wavelet space. The location of the ring coincides with the 
cold spot detected in 
Vielva et al. 2004 and Cruz et al. 2005.

\end{abstract}

%\begin{keywords}
%cosme ology: CMB -- data analysis
%\end{keywords}
 
\section{Introduction}

The Standard Inflationary model solves the horizon, the flatness, and the
monopole problems, and provides a framework for the formation of
structure in the universe (Guth 1981, Guth \& Pi 1982). Regarding the latter, the Standard Inflationary
model predicts the existence of quantum density fluctuations that are
amplified during the inflationary period and that grew, through
gravitational instabilities, into the galaxies and clusters that populate
our universe. These primordial density fluctuations are predicted 
to form a homogeneous and isotropic Gaussian field. The predicted statistical 
distribution of the 
Cosmic Microwave Background (CMB) temperature fluctuations reflects that 
of the primordial density fluctuations. Testing this prediction has 
been the aim of many works in the literature. In particular, since
the release of the first year of data collected by the WMAP (Wilkinson Microwave Anisotropy Probe) satellite (Bennett et al. 2003), a considerable number of papers have presented 
different statistical analyses based on data in real space, spherical harmonics, and
wavelet space (Park 2004, Eriksen et al. 2004a, Eriksen et al. 2004b, Eriksen et al. 2005, Larson \& Wandelt 2004, Hansen et al. 2004, Prunet et al. 2005, Vielva et al. 2004, Cruz et al. 2005, Mukherjee \& Wang 2004, Mc.Ewen et al. 2004). All these works have claimed the detection of deviations 
from the predictions of the Standard Inflationary model in the WMAP data set 
optimal for CMB studies (see Komatsu et al. 2003). 
Several other works have presented statistical
methods shown to be very powerful in detecting deviations from the 
Standard Inflationary model in the so-called Internal Linear Combination (ILC) map (Chiang et al. 2003, Coles et al. 2004, Copi et al. 2004, Naselsky et al. 2003, Chiang \& Naselsky 2004). In this case the most convincing source of deviations is foreground related. 

Deviations from the predictions of the 
Standard Inflationary model can have a cosmological origin. Non-Gaussianity 
can be generated under different conditions. A review on 
the predictions from several alternative scenarios including 
multi-field inflation, inhomogeneous reheating, non-linearities in the 
gravitational potential, and the curvaton-based model can be found in Bartolo et al. 2004.  However, these deviations can also be caused by systematic
effects or noise associated with the experiment as well as foregrounds (Galactic
or extra-galactic). The first year of data collected by the WMAP satellite 
has undergone careful characterization and examination by the WMAP team in an attempt to fully understand the data set (Page et al. 2003, Hinshaw et al. 2003, Bennett et al. 2003b, Barnes et al. 2003, Jarosik et al. 2003). The 
detection of deviations indicated in the previous paragraph show the
presence of some asymmetry between the northern and southern hemispheres. Moreover, analyses in wavelet space in different regions (north, south, northeast, northwest, southeast, southwest of the Galactic plane) performed by 
Vielva et al. 2004 and Cruz et al. 2005 indicate that 
the source of deviations might be a cold spot located at 
$(l\approx 209^{\circ}, b\approx -57^{\circ})$. The nature of the observed
deviations is still not clear. 

The implementation and development of new statistical methods is 
indispensable to improve our understanding of the deviations from 
the Standard Inflationary model predictions, in particular those 
observed in the WMAP data. There are many kinds of non-Gaussianity, and each type may be sensitive to some
statistical tests but immune to others. We need more types of tests, such as Higher Criticism in order to better understand non-Gaussianity. 
The Higher Criticism ($HC$) statistic was first  proposed in Donoho \& Jin (2004) for 
testing a very subtle problem: given $n$ Gaussian observations with same standard deviations but different means,  we want to test whether $n$  means are all zeros versus the alternative that a small fraction of them is nonzero. The fraction of nonzero means is too small to have any effect on the bulk of the observations,  so all statistics based on moments 
(e.g. excess kurtosis ($\kappa$)) would generally have no power for detection. With careful calibrations, $HC$ was proved to be optimal in detecting such situations. $HC$ is also competitive for detection in other scenarios. An analysis of simulated CMB on
flat patches of the sky using the $HC$ statistics 
was presented in Jin et al. 2004. 
The simulated images included CMB signals, as well as cosmic strings and 
Sunyaev-Zeldovich type emission. Analyses were performed in wavelet and curvelet spaces (Cand\'es \& Donoho 2000). The very particular non-Gaussianity introduced
by the simulated effects, as well as the restricted size of the sample, resulted in the lower sensitivity of this method in comparison with the 
traditional $\kappa$ statistic. However, theoretical studies indicate
that the $HC$ statistic will outperform the $\kappa$ when
the sample size gets larger. 

Besides its competitive detection power,  $HC$ is also valuable 
in identifying the origin of detected deviations, offering a 
way to determine which portion of the data contributes most to the deviation;  this property is not available for many other statistics such as $\kappa$.
As mentioned above, analyses of the WMAP 
data in wavelet space suggest that the source of the detected deviations
is due to the cold temperatures of certain pixels within a spot located
in the southern hemisphere. The $HC$ test is especially sensitive to abnormally high amounts of moderately large observations and, as shown in this paper, its application to WMAP data 
shows deviations from the predictions 
of the Standard Inflationary model. Moreover, inspection of
individual pixel $HC$ values in the WMAP data provides a direct
means to identify the pixels at the source of the deviation. No additional region-by-region analysis is needed. The computational cost is therefore strongly reduced, making it a promising 
test for analyzing future higher resolution data, such as data from the Planck mission\footnote[1]{http://www.rssd.esa.int/index.php?project=PLANCK}.

The paper is organized as follows. 
The $HC$ statistical test is described in
detail in Section 2. A comparison between the performance of 
this statistic and that of the kurtosis ($\kappa$) and the Maximum ($Max$) is included in subsection 2.1.
The WMAP data is analyzed in real and wavelet spaces. A description of the combination of WMAP data in different channels, the Monte Carlo simulations, and the process followed in wavelet
space is included in Section 3. The analysis 
of the WMAP data is presented in Section 4. 
Section 5 is dedicated to conclusions and discussion.

\section{Higher Criticism}

The $HC$ statistic was first proposed in Donoho \& Jin 2004, Jin 2004. 
Given $n$ independent observations of a distribution which is thought to be
slightly deviated from the standard Gaussian, one can compare the fraction
of observed significances at a given $\alpha$-level (i.e. the number 
of observed values exceeding the upper-$\alpha$ quantile of the standard Gausian)
  to the expected fraction
under the standard Gaussian assumption:
$$\sqrt n | (Fraction ~ of ~ Significance ~ at ~ \alpha )-\alpha| /(\alpha (1-\alpha ))^{0.5}.$$
The $HC$ statistic is then defined 
as the maximum of the above quantities over all significance levels $0<\alpha<1$.
Given $n$ individual observations $X_i$ from a distribution which is thought to 
be symmetric and slightly deviated from the standard Gaussian, there is a simpler equivalent form of HC defined as follows. First, we convert
the individual $X_i$'s  into individual $p$-values:  $p_i = P\{ |N(0,1)| > |X_i| \}$, 
then we let  $p_{(1)} < p_{(2)} < \ldots < p_{(n)}$ denote the $p$-values  sorted in 
ascending order,  define:
$$
HC_{n,i}  =    \sqrt{n} \biggl|  \frac{i/n  - p_{(i)}}{\sqrt{p_{(i)} (1-p_{(i)})}}  \biggr|,  
$$
the $HC$ statistic  is  then:
$$
HC_{n}^* =  \max_{i}  HC_{n,i}
$$
or in a modified form:
$$
HC_n^+  = \max_{\{i:  \; 1/n  \leq  p_{(i)} \leq  1 - 1/n \}} HC_{n,i};
$$
we let  $HC_n$ refer either to  $HC_n^*$  or $HC_n^+$ whenever there is no confusion.
The above definition is slightly different from that in Donoho \& Jin 2004, but  the ideas are essentially the same. 

$HC$ is useful in non-Gaussianity detection when $X_i$ are truly from $N(0,1)$, with the result that  $HC_{n,i} $ is approximately distributed as $N(0,1)$ for almost every $i$.   Thus an unusually large $HC_n^*$ or $HC_n^+$ value strongly implies non-Gaussianity.  Moreover,   $HC_{n,i}$ also provides {\it localized} information 
on deviations from Gaussianity. We can track down the source of non-Gaussianity  by studying which portion of 
the data gives unusually large  $HC_{n,i}$.

Previous works have claimed the detection of deviations from the predictions
of the Standard Inflationary model in the First Year WMAP data, using the 
$\kappa$ statistic in wavelet space (Vielva et al. 2004, Mukherjee \& Wang 2004, McEwen et al. 2004). In order to establish a comparison between 
this statistic and the $HC$ we will include calculations of $\kappa$ in our analysis as well. For completeness, we will also 
use the so-called $Max$ statistic.  The definitions for these
two statistics are provided below. A discussion about the theoretical 
power of the three statistics to detect deviations from Gaussianity is included in the following subsection.\hfill\break
\indent  $Kurtosis$, $\kappa$. For a (symmetric) random variable $X$, $\kappa$ is $\kappa(X) = \frac{E[X^4]}{(E[X^2])^2} -3$, which uses the 4th moment to measure the departure from Gaussianity.    
 $\kappa$ is useful in non-Gaussianity detection as $\kappa(X_1,X_2, \ldots,X_n) \approx \kappa(X)$ for large $n$.\hfill\break 
\indent $Max$.   The largest (absolute) observation is a classical statistic: 
$$
M_n   =  max\{ |X_1|, |X_2|, \ldots, |X_n|\}. 
$$
$Max$ is useful in non-Gaussianity detection because $M_n \approx \sqrt{2 \log n}$ when $X_i$ are truly from $N(0,1)$;  thus a significant difference between $M_n$ and $\sqrt{2 \log n}$ implies non-Gaussianity.

\subsection{Comparison of Higher Criticism, Max and Excess Kurtosis}

The aim of this section is to establish a simple 
theoretical comparison between the three statistics applied in this paper. 
We show the power of the 
different statistical tests in detecting the distortion generated by 
a faint non-Gaussian signal (modeled as a function of the decaying rate 
of the tail of the distribution) superposed on a Gaussian signal.   
Similar to the analysis presented in Jin et al. 2004, the superposed image can be thought of as $Y = N + G$, where $Y$ is the observed image, $N$ is the non-Gaussian component,  and $G$ is the Gaussian component (assumed to have mean zero and dispersion one, $N(0,1)$).  We study the power of the three statistics in testing whether $N = 0$ or not. 

One can do such a test either in real space (the space of the observations) or, as it is done in this paper, in wavelet space. 
For sufficiently fine resolution, the wavelet coefficients $X_i$ of $Y$ can be modeled as: 
$$
X_i = \sqrt{1 - \lambda }  \cdot z_i  + \sqrt{\lambda }  \cdot w_i,  ~~~~~~ 1 \leq i \leq n,  
$$   
where $n$ is the number of observations, $1\geq \lambda \geq  0$ is a parameter, $z_i \stackrel{iid}{\sim} N(0,1)$ are the transform coefficients
of the Gaussian component $G$, and $w_i \stackrel{iid}{\sim} W$ are the transform coefficients of the non-Gaussian component $N$. $W$ is some unknown distribution with vanishing first and third order moment 
(this constraint is only imposed to 
simplify the possible range of non-Gaussian distributions, and it will be
representative of a Gaussian distribution modified by a few large values contributing to a tail). Without loss of generality, we assume the standard deviation of both $z_i$ and $w_i$ are $1$.  Phrased in statistical terms, the problem of detecting the existence of a non-Gaussian component is equivalent to discriminating between the null hypothesis and the alternative hypothesis:  
%\begin{align}
$$H_0: \;\;\;   X_i = z_i,  \label{EqHypo1}   $$
$$H_1:   X_i = \sqrt{1 - \lambda } \cdot z_i  + \sqrt{\lambda } \cdot  w_i,   \qquad 0 \le \lambda \le 1.    \label{EqHypo2}$$
%\end{align}
\noindent $N \equiv 0$ being  equivalent to $\lambda \equiv 0$.  

In order to obtain a quantitative estimation of the ability of the three
statistics to detect the non-Gaussian component, we parametrize the tail probability of $W$ as follows.
$$
\lim_{x \rightarrow \infty}   x^{\alpha}  P\{|W|  > x\}   = C_{\alpha},  ~~~~~~ \mbox{$C_{\alpha}$ is a constant. }  
$$
To model increasingly challenging situations as the number of observations increases, we calibrate $\lambda $ to decay with $n$ as:
$$
\lambda    =   \lambda _n   =  n^{-r}.  
$$
\noindent To constrain the detectability of such a non-Gaussian distribution superposed to a Gaussian one, we search the $r$ and $\alpha$ parameter space. We define the regions of this space that will be detectable
under different statistical tests. It was shown in Jin et al. 2004 that there is a curve in the 
$r$-$\alpha$ plane that separates the detectable regions of parameters from the undetectable regions.
That curve is given by:
$$
r=   \left\{ 
\begin{array}{ll}
2/\alpha,   &\qquad \alpha \leq 8, \\
1/4,  &\qquad  \alpha > 8. 
\end{array}
\right.
$$
In Figure 1, we compare the results of $HC$, $Max$ and $\kappa$.  When it is possible to detect,  $HC$ or $Max$ are better than $\kappa$ when $\alpha \leq 8$.  $\kappa$ is better than $HC$ 
or $Max$ when $\alpha > 8$.

To conclude this section, we remark that the performance of $HC$, $Max$, and $\kappa$ depends on different situations of non-Gaussianity. 
Intuitively, $Max$ is designed to capture evidence of unusual behavior of the most extreme observations against the Gaussian assumption. The $HC$ statistic is able to capture unusual behavior of the most extreme observations, as well as unusually large amounts of moderately high observations. Thus $HC$ is better than $Max$, in general. However, when the evidence against the Gaussian assumption truly lies in the most extreme observations, $HC$ and $Max$ are almost equivalent.  In contrast, $\kappa$ is designed to capture the evidence hidden in the 4th moment. Therefore this statistic depends on the bulk of the data, rather than on a few extreme observations or a small fraction of relatively large observations. 
Lastly, the $HC$ statistic offers an automatic way to find the area of the data accounting for the non-Gaussianity, while the $Max$ and $\kappa$ statistics do not have this capability.

\setcounter{figure}{0}
\begin{figure*}
 \epsfxsize=84mm
 \epsffile{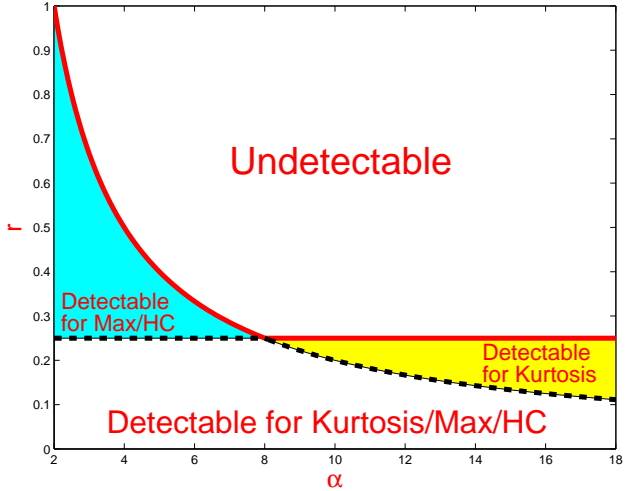}
 \caption{Detectable regions  in the $\alpha-r$ plane.  With $(\alpha,r)$ in the white region on the top or the undetectable region, all possible statistics fail asymptotically for detection. With $(\alpha,r)$ in the white region on the bottom,  both $\kappa$ and $Max/HC$ are able to detect reliably.      While in the blue region to the left,  $Max/HC$ is able to detect reliably, but $\kappa$ completely fails. In the yellow region to the right, $\kappa$ is able to detect reliably, but $Max/HC$  completely fail asymptotically.}
 \label{f1}
\end{figure*}

\section{WMAP First Year Data and Simulations}

The data collected by the WMAP satellite during the first year of operation 
is available at the Legacy Archive for Microwave Background Data Analysis 
(LAMBDA) website\footnote[2]{http://lambda.gsfc.nasa.gov/}. Following 
Komatsu et al. 2003, the analysis presented in this work is based on 
a data set obtained by the weighted combination of the released Foreground Cleaned Intensity Maps at bands Q,V and W. Each of these maps can be downloaded 
in the HEALpix\footnote[3]{HEALPix http://www.eso.org/science/healpix/} nested format at resolution $nside=512$ (total number
of pixels being $12\times nside^2$). The co-added map is obtained by the 
following combination
$$T(i)={{\sum_{r=3}^{10} T_r(i)w_r(i)}\over {\sum_{r=3}^{10} w_r(i)}}.$$
The temperature at pixel $i$, $T(i)$ results from the ratio of the weighted sum 
of temperatures at pixel $i$ at each radiometer divided by the sum of the 
weights of each radiometer at pixel $i$. The radiometers Q1, Q2, V1, V2, W1, W2, W3 and W4 are sequentially numbered from 3 to 10. The weights at each pixel,
for each radiometer $w_r(i)$, are the ratio of the number of observations
$N_r(i)$ divided by the square of the receiver noise dispersion $\sigma_{o,r}$.This results in a co-added map at resolution $nside=512$. This map 
is downgraded to resolution $nside=256$ before the analyses are performed. 
In order to remove Galactic and point source emission, we applied the 
most conservative mask provided by the WMAP team, the Kp0 mask. 
This follows the procedure first presented by Vielva et al. 2004.

In order to detect any possible deviations from the predictions of 
the Standard Inflationary model we compared the values of several statistics 
in the WMAP data set described above, with those obtained from
5000 Monte Carlo simulations. The 
temperature at a certain pixel $i$ (pointing towards a direction characterized by polar angles $\theta_i$ and $\phi_i$), can be expressed as an expansion
in spherical harmonics $Y_{lm}(\theta_i,\phi_i)$
$$T(\theta_i,\phi_i)=\sum_{l,m}a_{lm}Y_{lm}(\theta_i,\phi_i)$$
The Monte Carlo simulations were performed assuming
a Gaussian distribution $N(0,C_l)$ for the spherical harmonic coefficients $a_{lm}$ where $C_l$ represents the power spectrum that best fits the WMAP, CBI and ACBAR CMB data, plus the 2dF and Lyman-alpha data. Beam transfer functions as well as number of observations and receiver noise dispersion were taken
into account when simulating data taken by each of the receivers (all of these, as well as the best fit power spectrum are provided by the WMAP team in
the LAMBDA website). 

The analysis was carried out in both real space and wavelet
space. We convolved the WMAP data set and the Monte Carlo simulations with 
the Spherical Mexican Hat Wavelet (SMHW) at fifteen scales, following the
same procedure presented in Vielva et al. 2004 and Cruz et al. 2005. The scales,
numbered from 1 to 15, correspond to $13.74$, $25.0$, $50.0$, $75.0$, $100.0$, $150.0$, $200.0$, $250.0$, $300.0$, $400.0$, $500.0$, $600.0$, $750.0$, $900.0$, and $1050.0$ arcmin. 
In order to avoid border effects caused by the mask included 
in the WMAP data set and the simulations, analyses are performed outside 
extended masks defined at each scale. Given a scale, the extended mask 
is the Kp0 mask plus pixels near the Galactic plane that are 
within $2.5$ times the scale. All of these extended masks were presented in 
Figure 2 of Vielva et al. 2004. As discussed in the next section, deviations 
are detected in wavelet space.  This shows once again the value of wavelets to 
provide a space in which certain features that might be causing
deviations from the predictions of the Standard Inflationary model are enhanced.

\section{ANALYSIS}

Based on 5,000 simulations, 
we calculated the $68\%$, $95\%$, and $99\%$ confidence regions for each of the 4
statistics ($\kappa$, $Max$, and the two HC tests) at each of the 15 wavelet scales used.  We used two-sided
confidence regions for $\kappa$, as it is symmetric about 0 under the null hypothesis that the data is Gaussian.
$Max$ and $HC$ statistics are defined so that the larger the statistic, the stronger the evidence against
the null hypothesis. Therefore, we used one-sided confidence regions for these statistics.

Figure $2$ shows the results obtained based on $\kappa$. Dots denote
the value of this statistic for the WMAP data set. Bands outlined by dashed,
dotted-dashed,  and solid lines correspond to the $68\%$, $95\%$, and $99\%$
confidence regions respectively (symbols and lines will represent the same in the
figures included in this paper unless a comment is added). Clearly, the
results agree with those presented in Vielva et al. 2004. 
Figures $3$ and $4$ show the new
results based on the $Max$ and $HC$ statistical tests. The WMAP data
appears outside the $99\%$ confidence level obtained based on the predictions of the Standard
Inflationary model at scale 9 (300 arcmin) for these three tests. In particular, $\approx 99.46\%$ of the 5000 simulations have $Max$ and $HC/HC^{+}$ 
values below the 
one obtained from the WMAP data set. Therefore in this particular case these statistics are as competitive as the kurtosis in detecting deviations from 
the assumed null hypothesis.

\setcounter{figure}{1}
\begin{figure*}
 \epsfxsize=84mm
 \epsffile{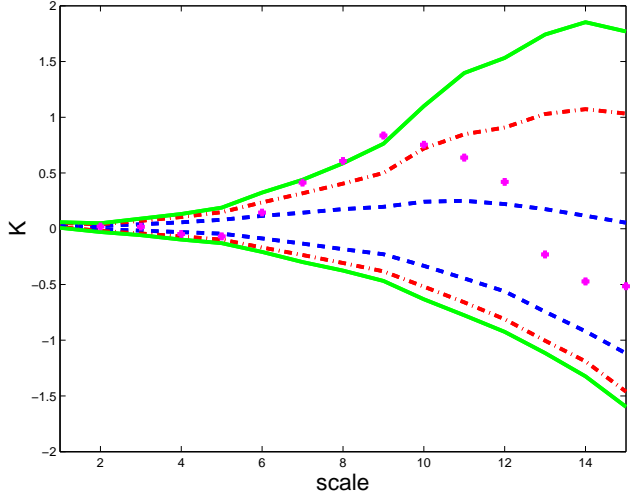}
 \caption{Values of the $\kappa$ statistic for the analyzed WMAP data
set (dots). The bands outlined
by dashed, dotted-dashed and solid lines correspond to the $68\%$, $95\%$ and $99\%$ confidence regions respectively.}
 \label{f1}
\end{figure*}

\setcounter{figure}{2}
\begin{figure*}
 \epsfxsize=84mm
 \epsffile{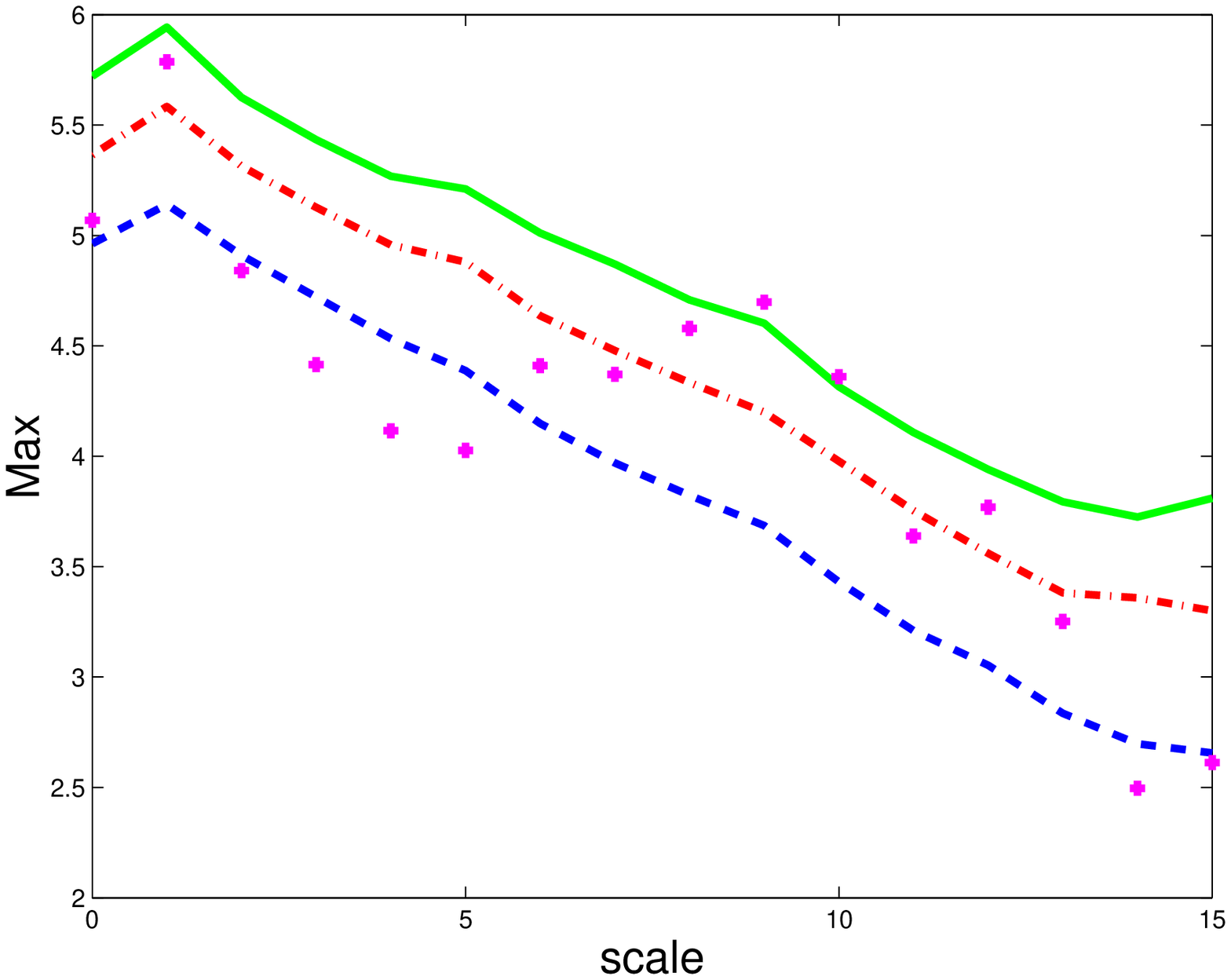}
 \caption{Values of the $Max$ statistical test for the analyzed WMAP data
set (dots). The dashed, dotted-dashed and solid lines correspond to the $68\%$, $95\%$ and $99\%$ confidence levels respectively.}
 \label{f1}
\end{figure*}

\setcounter{figure}{3}
\begin{figure*}
 \epsfxsize=84mm
 \epsffile{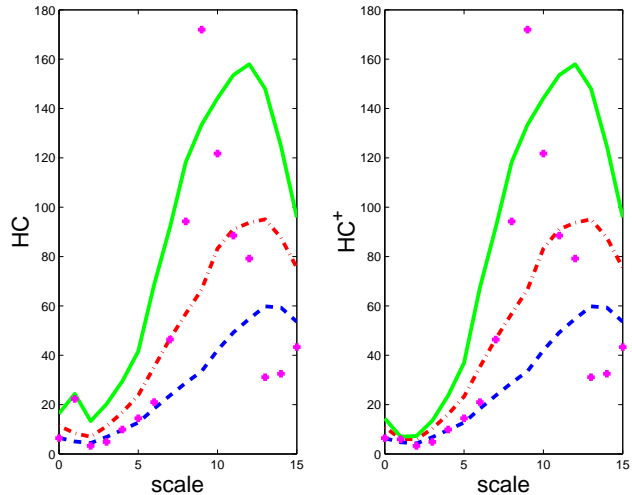}
 \caption{Values of the $HC/HC^{+}$ statistical tests for the analyzed WMAP data
set (dots). The dashed, dotted-dashed and solid lines correspond to the $68\%$, $95\%$ and $99\%$ confidence levels respectively.}
 \label{f1}
\end{figure*}

\subsection{Location of the outlying pixels provided by the Higher Criticism statistic} 

As discussed above, the $HC$ statistical tests introduced in this paper are able to
detect deviations from the Standard Inflationary model in the WMAP data at scales 
around 300 arcmin. Moreover, the power of this new test resides in 
providing a direct way to determine which pixels in the WMAP data set
are generating these deviations. By comparing the values of the $HC$ test at each pixel, with the 
value that defines the $99\%$ confidence limit
at scale 9, we were able to extract a total of $490$ pixels that are causing
the detected deviation. Figure 5 shows the individual pixel values of the
$HC$ statistic for the WMAP data set at scale 9. The selected
pixels above the $99\%$ limit are plotted 
in the map shown in Figure 6. As can be seen, these pixels define 
a ring 
centered at position $(l\approx 209^{\circ},b\approx -57^{\circ})$. 
It is important to note that the correlations introduced
by the convolution with the wavelet need to be taken into account in order to 
properly interpret this result. Some pixels within the ring could also initially 
(in the map in real space) deviate from Gaussianity.

\subsection{Effects of the subtraction of pixels at the source of the detection}

Removal of the pixels in the ring from the analysed WMAP data resulted in a set of data
compatible with the predictions from the Standard Inflationary model (see Figure 7). The deviation 
observed in the $\kappa$ statistic at the wavelet scale of $5$ degrees, in 
this paper as well as in previous papers by Vielva et al. 2004, Cruz et al. 2005,
Mukherjee \& Wang 2004, McEwen et al. 2004, 
disappeared as the pixels in the ring $(l\approx 209^{\circ},b\approx -57^{\circ})$ were removed. These pixels are part of the 
cold spot pointed out by Vielva et al. 2004 and Cruz et al. 2005 as
being the source of the observed deviations. The $Max$ and the $HC$ values decrease as well (after removal of the 
pixels in the ring) fitting now within the $68\%$ confidence
region. 
The Higher Criticism statistic is useful by offering an automatic 
detection of the non-Gaussian 
portions of the data.  This characteristic is 
not easily available for other tests, such as Kurtosis.

\setcounter{figure}{4}
\begin{figure*}
\epsfxsize=84mm
 \epsffile{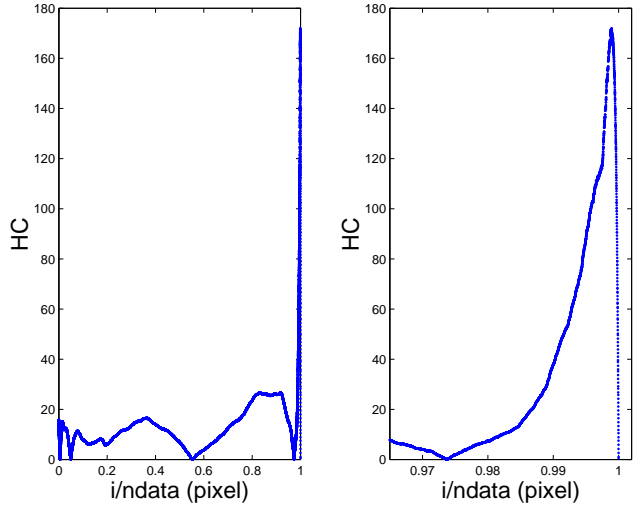}
 \caption{Values of the $HC$ statistical test at every pixel of the analyzed WMAP data
set. After sorting all the p-values, a number is assigned 
to each pixel $i$. That number goes from 0 (pixel with the smallest p-value) to the 
number of data (pixel with the largest p-value). The $HC$ values at each pixel
are plotted against $i/number~of~data$. The region corresponding to 
the extreme values of $HC$ is shown in the figure on the right panel.}
 \label{f1}
\end{figure*}

\setcounter{figure}{5}
\begin{figure*}
 \epsfxsize=100mm
 \epsffile{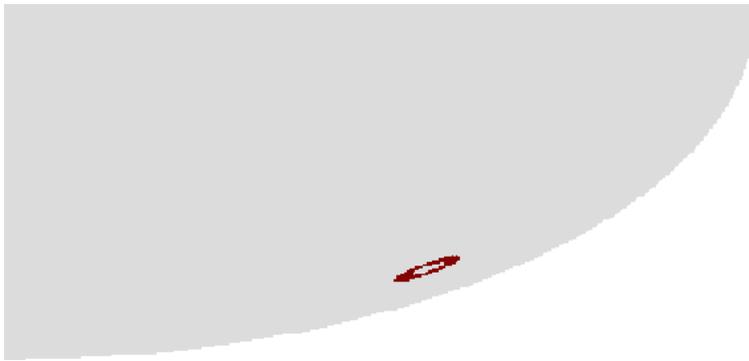}
 \caption{A quarter of the whole sky showing the 
location of the WMAP pixels with $HC$ values above the extreme of the $99\%$ limit obtained from 5000 simulations.}
 \label{f1}
\end{figure*}

\setcounter{figure}{6}
\begin{figure*}
 \epsfxsize=100mm
 \epsffile{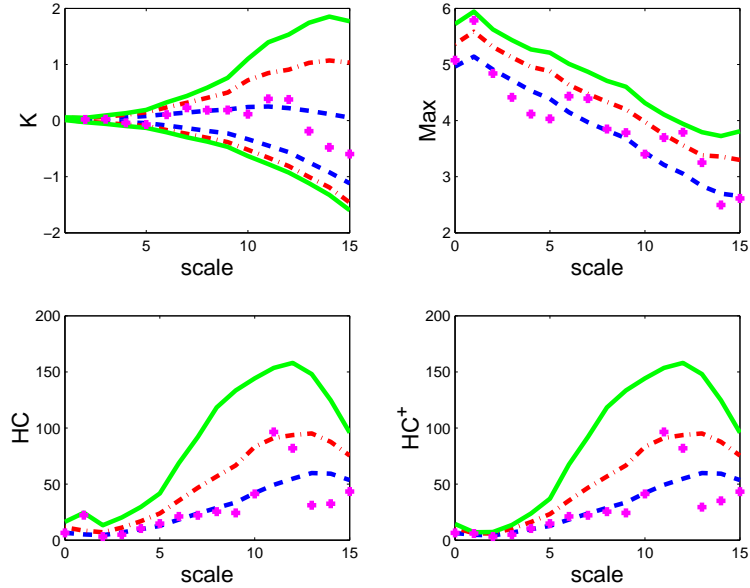}
 \caption{Values of $\kappa$, $Max$ and $HC$ tests for the analyzed WMAP data
set after subtracting the pixels that were causing the deviations from 
the predictions of the Standard Inflationary model (dots). The bands outlined
by dashed, dotted-dashed and solid lines correspond to the $68\%$, $95\%$ and $99\%$ confidence regions respectively for $\kappa$. The dashed, dotted-dashed and solid lines correspond to the $68\%$, $95\%$ and $99\%$ confidence levels respectively for $Max$, $HC$ and $HC^{+}$. }
 \label{f1}
\end{figure*}

\section{Conclusions and Discussion}

This paper presents an analysis of the compatibility of the distribution of 
the CMB temperature fluctuations observed by WMAP with the predictions from
Standard Inflation. The analysis is based on the recently developed
$HC$ statistic. 
This statistic is especially sensitive to two types of deviations from Gaussianity: an unusually large amount of moderately significant values or a small fraction of unusually extreme values. In both of these cases the HC statistic is optimal (Donoho \& Jin 2004, Jin et al. 2004). Moreover, the definition of the statistic naturally suggests a direct way to locate the possible sources of non-Gaussianity.
%The power of this statistic resides on its ability 
%to detect unusual behavior caused by very extreme values as well as by 
%a considerable amount of moderately large values. Moreover, it provides a 
%direct way to identify the source of any possibly observed unusual behavior. 
%The definition of the statistic used in this work is optimal for
%detecting deviations from a Gaussian distribution. 
Even if the wavelet
transform is a linear process, the original distribution is slightly distorted as
the scale of the wavelet, and the area covered by the mask, increase. 
We are working on improving the 
power of the $HC$ statistic to be optimal in the detection
of deviations from Gaussianity when applied in wavelet space.  

We compared the performance of the $HC$ statistic with that
of $\kappa$ and $Max$. The comparison was based on the 
analysis of the first year WMAP data, in real space and in wavelet space. 
Distributions of the three statistics were built on 5000 simulations assuming the Standard
Inflationary model predictions as well as the constraints coming from the 
WMAP observations. The three statistics provided comparable results, pointing 
to the presence of deviations at a wavelet scale of $5$ degrees. We made use
of the $HC$ statistic to automatically identify the pixels in the WMAP data set that were
causing such deviations. A ring centered at $(l\approx 209^{\circ}, b\approx -57^{\circ})$ containing $490$ pixels was shown to be the cause behind the 
detected deviations. Removal of this set of pixels from the WMAP data made the data compatible with the predictions from the Standard Inflationary 
model. One should be cautious when interpreting this result.      
The detection is achieved 
in wavelet space. Even if only the pixels in the ring appear as outliers of the HC 
distribution at wavelet scale of 5 degrees, some other pixels whithin the ring 
could be involved in the observed deviation. 
Convolution with the wavelet introduces correlations between pixels that need to be
properly taken into account in the definition of the statistic to allow a correct
interpretation. 
It is important to note that the pixels selected as the source of the observed deviations are part of the
cold spot pointed out by Vielva et al. 2004 and Cruz et al. 2005. A careful study of the possible influence of foregrounds, noise, and beam distortion and assumption of a certain power spectrum was carried out in those two papers. 

To conclude we would like to remark on the power of the $HC$ statistic for 
detecting and locating possible sources of non-Gaussianity. In particular, regarding analysis of the
CMB, the $HC$ statistic can be useful at several steps in the 
data processing and final analysis, from the study of distortions 
caused by systematic effects in the time-ordered data to the study of 
compatibility of the statistical distribution of the observed temperature fluctuations with predictions from theories of the very early universe.

\section*{Acknowledgements}

We thank P. Vielva for providing the extended masks used at different
wavelet scales. We would like to thank Douglas Crabill for helping with
some of the figures.
We acknowledge the use of the Legacy Archive for Microwave Background Data Analysis (LAMBDA). Support for LAMBDA is provided by the NASA Office of Space
Science. 
Some of the results in this paper have been derived using the HEALPix (G\'orski, Hivon, and Wandelt 1999) package.
L.C. and A.T. acknowledge the Indiana Space Grant Consortium for financial support. A.T. acknowledges the Spira Fellowship (Summer 2004 and Summer 2005).


\begin{thebibliography}{}


\bibitem{} Barnes, C. et.al., 2003, ApJS, 148, 51


\bibitem{} Bartolo, N., Komatsu, E., Materrese, S. \& Riotto, A. 2004, Phys. Rept. 402, 103 


\bibitem{} Bennett, C.L. et al. 2003, ApJS, 148, 1

\bibitem{} Bennett, C.L. et.al., 2003b, ApJS, 148, 97

\bibitem{} Cand\'es, E.J. \& Donoho, D.L. 2000, In {\it Curve and Surface  Fitting: Saint-Malo 1999}, Nashville, TN. Editors: Cohen, A., Rabut, C. \& Schumaker, L.L.. Vanderbilt University Press, pag. 105 


\bibitem{} Chiang, L.-Y., Naselsky, P.D., Verkhodanov, O.V. \& Way, M.J. 2003, ApJ, 590, L65

\bibitem{} Chiang, L.Y. \& Naselsky, P.D. 2004, astro-ph/0407395

\bibitem{} Coles, P., Dineen, P., Earl, J. \& Wright, D. 2004, MNRAS, 350, 989

\bibitem{} Copi, C.J., Huterer, D. \& Starkman, G.D. 2004, Phys.Rev.D, 70, 043515

\bibitem{} Cruz, M., Mart\'\i nez-Gonz\'alez, E., Vielva, P. \& Cay\'on, L. 2005, 356, 29


\bibitem{} Donoho, D.  \& Jin, J.  2004, {\it Ann. Statist.}, {\bf 32}, 962


\bibitem{} Eriksen, H.K., Novikov, D.I., Lilje, P.B., Banday, A.J. \& G\'orski, K.M. 2004a, ApJ, 612, 64

\bibitem{} Eriksen, H.K., Hansen, F.K., Banday, A.J., G\'orski, K.M. \& Lilje, P.B. 2004b, ApJ, 605, 14

\bibitem{} Eriksen, H.K., Banday, A.J., G\'orski, K.M. \& Lilje, P.B. 2005, ApJ, 622, 58

\bibitem{} G\'orski, K.M., Hivon, E. \& Wandelt, B.D. 1999, in Proceedings of the MPA/ESO Cosmology Conference "Evolution of Large-Scale Structure", eds. A.J. Banday, R.S. Sheth and L. Da Costa, PrintPartners Ipskamp, NL, pp. 37-42 (also astro-ph/9812350)

\bibitem{} Guth, A.H. 1981, Phys.Rev.D., 23, 347

\bibitem{} Guth, A.H. \& Pi, S.-Y. 1982, Phys.Rev.Lett., 49, 1110

\bibitem{} Hansen, F.K., Banday, A.J. \& G\'orski, K.M. 2004, astro-ph/0404206

\bibitem{} Hinshaw, G. et.al., 2003, ApJS, 148, 63


\bibitem{} Jarosik, N. et.al., 2003, ApJS, 148, 29

\bibitem{} Jin, J. 2004, Institue of Mathematical Statistics Monograph, No. 45, 255

\bibitem{} Jin, J., Starck, J.-L., Donoho, D.L., Aghanim, N. \& Forni, O. 2004, submitted to EURASIP Journal on Applied Signal Processing, special issue on ``Applications of Signal Processing in Astrophysics and Cosmology''



\bibitem{} Komatsu, E. et al. 2003, ApJS, 148, 119  


\bibitem{} Larson, D.L. \& Wandelt, B.D. 2004, ApJ, 613, L85

\bibitem{} McEwen, J.D., Hobson, M.P., Lasenby, A.N. \& Mortlock, D.J. 2005, MNRAS, 359, 1583

\bibitem{} Mukherjee, P. \& Wang, Y. 2004, ApJ, 613, 51

\bibitem{} Naselsky, P.D., Doroshkevich, A.G. \& Verkhodanov, O.V. 2003, ApJ, 599, L53

\bibitem{} Page, L. et al. 2003, ApJS, 148, 39  

\bibitem{} Park, C.-G. 2004, MNRAS, 349, 313

\bibitem{} Prunet, S., Uzan, J.-P., Bernardeau, F. \& Brunier, T. 2005, Phys. Rev. D, 71, 083508 

\bibitem{} Vielva, P., Mart\'\i nez-Gonz\'alez, E., Barreiro, R.B., Sanz, J.L. \& Cay\'on, L. 2004, ApJ, 609, 22

\end{thebibliography}
\end{document}